\documentclass[prd,showpacs,preprintnumbers,amsmath,amssymb]{revtex4}

\usepackage{graphicx}
\usepackage{dcolumn}
\usepackage{bm}
\def\be{\begin{eqnarray}}
\def\ee{\end{eqnarray}}
\def\bea{\begin{eqnarray}}
\def\eea{\end{eqnarray}}

\def\0T{{\bf 0}_\perp}
\begin{document}


\title{Quark Orbital Angular Momentum in the MIT Bag Model}

\author{Matthias Burkardt and Abdullah Jarrah}
 \affiliation{Department of Physics, New Mexico State University,
Las Cruces, NM 88003-0001, U.S.A.}

\begin{abstract}
Using the MIT bag model, we study the contribution from the gluon 
vector potential due to the spectators to the
orbital angular momentum of a quark in the bag model. For 
$\alpha_s = {\cal O}(1)$, this spectator contribution to the quark orbital 
angular momentum in the gauge-covariant Ji decomposition is of the
same order as the non gauge-covariant quark orbital angular momentum
and its magnitude is larger for $d$ than for $u$ quarks and negative 
for both.
\end{abstract}

\maketitle
\narrowtext
\section{Introduction}
Since the famous EMC experiments revealed that only a small fraction
of the nucleon spin is due to quark spins \cite{EMC}, 
there has been a great
interest in `solving the spin puzzle', i.e. in decomposing the
nucleon spin into contributions from quark/gluon spin and
orbital degrees of freedom.
In this effort, the Ji decomposition \cite{JiPRL}
\be
\frac{1}{2}=\frac{1}{2}\sum_q\Delta q + \sum_q { L}_q^z+
J_g^z
\label{eq:JJi}
\ee
appears to be very useful, 
as not only the quark spin contributions $\Delta q$ but also
the quark total angular momenta $J_q \equiv \frac{1}{2}\Delta q + 
{ L}_q^z$ (and by subtracting the spin piece also the
the quark orbital angular momenta $L_q^z$) entering this decomposition
can be accessed experimentally, through generalized parton 
distributions (GPDs).
The terms in (\ref{eq:JJi}) are defined as expectation
values of the corresponding terms in the angular momentum tensor
\be
M^{0xy}= \sum_q \frac{1}{2}q^\dagger \Sigma^zq +
\sum_q q^\dagger \left({\vec r} \times i{\vec D}
\right)^zq
+  
\left[{\vec r} \times \left({\vec E} \times {\vec B}\right)\right]^z
\label{M012}
\ee
in a nucleon state with zero momentum. Here
$i{\vec D}=i{\vec \partial}-g{\vec A}$ is the gauge-covariant
derivative.
The main advantages of this decomposition are that each term can be 
expressed as the
expectation value of a manifestly gauge invariant
local operator and that the
quark total angular momentum $J_q^z=\frac{1}{2}\Delta q+L_q^z$
can be related to generalized parton distributions (GPDs) 
\cite{JiPRL} 
and is thus accessible in deeply virtual Compton scattering and
deeply virtual meson production and can also be
calculated in lattice gauge theory.

Recent lattice calculations of GPDs surprised in several ways
\cite{lattice}.
First, the light quark orbital angular momentum (OAM) is consistent 
with
$L_u\approx -L_d$, i.e. $L_u+L_d\approx 0$, which
would imply that $J_g \approx \frac{1}{2}\cdot 0.7$ represents
the largest piece in the nucleon spin decomposition. Secondly,
$L_u\approx -0.15$ and $L_u\approx +0.15$ in these calculations, 
i.e. the
opposite signs from what one would expect from many quark
models with relativistic effects, as we will also illustrate in the
following section. While the inclusion of still-omitted
disconnected diagrams may change the sum $L_u+L_d$, it does not
affect the difference $L_u-L_d$. In Ref. \cite{Tony}, it was
pointed out that evolution from a quark model scale of few hundred
MeV to the lattice scale of few GeV could possibly account for the
difference. However, due to the presence of interactions  
 through the vector potential in the gauge covariant derivative
$L_q^z$ does not have a parton interpretation, which
complicates its physical interpretation.

Jaffe and Manohar have proposed an alternative decomposition of the
nucleon spin, which does have a partonic interpretation
\cite{JM}
\be
\frac{1}{2}=\frac{1}{2}\sum_q\Delta q + \sum_q {\cal L}_q^z+
\frac{1}{2}\Delta G + {\cal L}_g^z,
\label{eq:JJM}
\ee
and whose terms are defined as matrix elements of the corresponding
terms in the $+12$ component of the angular momentum tensor
\be
M^{+12} = \frac{1}{2}\sum_q q^\dagger_+ \gamma_5 q_+ +
\sum_q q^\dagger_+\left({\vec r}\times i{\vec \partial}
\right)^z q_+  
+ \varepsilon^{+-ij}\mbox{Tr}F^{+i}A^j
+ 2 \mbox{Tr} F^{+j}\left({\vec r}\times i{\vec \partial} 
\right)^z A^j.
\label{M+12}
\ee
The first and third term in (\ref{eq:JJM}),(\ref{M+12}) are the
`intrinsic' contributions (no factor of ${\vec r}\times $) 
to the nucleon's angular momentum $J^z=+\frac{1}{2}$ and have a 
physical interpretation as quark and gluon spin respectively, while
the second and fourth term can be identified with the quark/gluon
OAM.
Here $q_+ \equiv \frac{1}{2} \gamma^-\gamma^+ q$ is the dynamical
component of the quark field operators, and light-cone gauge
$A^+\equiv A^0+A^z=0$ is implied. 
The residual gauge invariance can be fixed by
imposing anti-periodic boundary conditions 
${\bf A}_\perp({\bf x}_\perp,\infty^-)=-
{\bf A}_\perp({\bf x}_\perp,-\infty^-)$ on the transverse components
of the vector potential.

Only the $\Delta q$ are common to both decompositions. While for
a nucleon at rest the difference in the Dirac structure between
$L_q^z$ and ${\cal L}_q^z$ plays no role \cite{BC}, 
the appearance of the
gluon vector potential in the operator defining
$L_q^z$ implies that in general ${\cal L}_q^z\neq L_q^z$.
Indeed, in Ref. \cite{BC} their difference was illustrated in the
context of an electron in QED to order $\alpha$.
This lowest order correction to the electron orbital angular momentum
from its own vector potential, can be directly translated into
a QCD correction to the orbital angular momentum of a quark,
resulting in \cite{BC}
\be
L_q^z = {\cal L}_q^z - \frac{\alpha_s}{3\pi}
\label{eq:BC}
\ee
for a quark state with $J^z=+\frac{1}{2}$ (and ${\vec p}=0$).
However, in addition to this contribution to 
$q^\dagger \left({\vec r}\times g{\vec A}\right)q$ from the ${\vec A}$ field
`caused' by its own current, the vector potential can also have
been caused by the spectators, or may be due to intrinsic glue.
In this note, we use the bag model to make an estimate for
such `spectator effects', by examining the contribution of the vector
potential arising from spectator currents to $L_q^z - {\cal L}_q^z$
in perturbation theory to ${\cal O}(\alpha_s)$.

\section{Orbital Angular Momentum in the MIT Bag Model}
In the bag model \cite{MIT}, the Dirac wave function 
for a quark state with $j^z=s$, is of the form
\be
\psi_s = {\cal N}\left( \begin{array}{c} j_0(kr)\chi_s\\
ij_1(kr) \hat{\vec r}\cdot {\vec \sigma} \chi_s \end{array}\right)
\ee
where $\chi_s$ is a two-component Pauli spinor, $k=E=\frac{2.0428}{R}$,
with $R$ being the bag radius, and ${\cal N}={\cal N}_s$ is
a normalization constant. 
For simplicity, we take the quarks to be massless.

Evaluating the orbital angular momentum is straightforward
in the absence of QCD corrections. For a state with $s=+\frac{1}{2}$
one finds from the lower component of the Dirac wave function
\be
{\cal L}_q^z = \left|{\cal N}\right|^2\frac{8\pi}{3}\int_0^Rdr
r^2 j_1^2(kR) = 0.1735
\label{eq:bagLq}
\ee
One can easily understand the sign of (\ref{eq:bagLq}): since the
quark spin $s^z=\pm\frac{1}{2}$
and its orbital angular momentum must add up to
$j^z=\frac{1}{2}$ for each wave function component, 
the $\hat{z}$-component of the quark
orbital angular momentum can only be +1 or 0, i.e. 
$0\leq {\cal L}_q^z \leq 1$. For a quark state with 
$j_q^z=-\frac{1}{2}$, one obtains of course the opposite sign for
${\cal L}_q^z$.

Using standard $SU(6)$ wave functions, 
\bea
\left|p\uparrow\right\rangle = \frac{1}{\sqrt{18}}
& &\!\!\!\!\left\{2\left|u\uparrow u\uparrow d\downarrow \right\rangle
+2\left|u\uparrow d\downarrow u\uparrow\right\rangle
+2\left|d\downarrow u\uparrow u\uparrow \right\rangle
-\left|u\uparrow u\downarrow d\uparrow \right\rangle
-\left|u\downarrow u\uparrow d\uparrow \right\rangle\right.
\label{eq:SU6}\\
& &\quad 
-\left.\left|u\uparrow d\uparrow u\downarrow \right\rangle
-\left|u\downarrow d\uparrow u\uparrow \right\rangle
-\left|d\uparrow u\uparrow u\downarrow \right\rangle
-\left|d\uparrow u\downarrow u\uparrow \right\rangle\right\},
\nonumber
\eea
this result gets multiplied by
$\frac{4}{3}$ for $u$ quarks in a proton and by $-\frac{1}{3}$ for 
$d$ quarks in a proton,
yielding
\bea
{\cal L}^z_{u/p} &=& \frac{4}{3}\cdot 0.1735 = 0.2313 \label{eq:num1}\\
{\cal L}^z_{d/p} &=& -\frac{1}{3}\cdot 0.1735 = -0.0578, \nonumber
\eea
with signs opposite to that of the lattice calculation \cite{lattice}.
A similar pattern for the signs as in Eq. (\ref{eq:num1})
is observed in all quark models where the orbital angular momentum
arises as a relativistic effect from the lower Dirac component.
Similarly, phenomenological models for single-spin asymmetries have
the same pattern of signs for quark angular momenta as 
the bag model (\ref{eq:num1}) and thus also differ from the
lattice QCD results \cite{lattice}. Although present lattice
calculations still suffer from uncontrolled systematic errors due
to the omission of operator insertions into disconnected quark loops,
this does not affect the isovector combination ${\cal L}^z_{u/p}-
{\cal L}^z_{d/p}$, which is positive in the model calculations but
negative in lattice QCD.
$Q^2$ evolution has been proposed as a possible explanation to account
for this apparent discrepancy \cite{Tony,Wakamatsu}, 
but in order to accomplish
quantitative consistency, perturbative evolution equations
need to be used very deep in the nonperturbative regime.

Since quark models have, to lowest order in $\alpha_s$, no vector 
potential, it makes perhaps more sense to identify
the quark OAM from these models with ${\cal L}^z_q$ rather than with
the GPD-based $L^z_q$. In Ref. \cite{BC} the difference 
$\Delta^z_e \equiv {\cal L}_e^z-L_e^z$ was calculated to order
$\alpha$ for a single electron in QED and the result then also
applied to a single quark in QCD. However, in QCD quarks are never 
alone and the question arises regarding the effects from
`spectator currents' on the orbital angular momentum of each quark.

In order to address this issue,  we will
in the following focus on estimating ${\cal O}(\alpha_s)$
corrections to the difference
\be
\Delta^z_q \equiv L_q^z - {\cal L}_q^z= 
\langle q^\dagger\left({\vec r} \times g{\vec A}\right)^zq\rangle .
\label{eq:Delta}
\ee
The vector potential in (\ref{eq:Delta}) is calculated from
the spectator currents, which are obtained by taking matrix elements
in the corresponding ground state bag model wave functions.
The vector potential resulting from these static currents
is obtained by solving
\be
{\vec \nabla}^2 {\vec A}^{a}({\vec r}) = - {\vec j}^{a}({\vec r})
= - \sum_{s^\prime} g\psi^\dagger_{s^\prime}({\vec r})
{\vec \alpha} \frac{\lambda^a}{2} \psi_{s^\prime}({\vec r})
\ee
for each color component $a$ and
where the summation is over the spectators
(here we pick the gauge ${\vec \nabla}\cdot {\vec A}^a=0$, but 
to ${\cal O}(\alpha_s)$ the
result is actually gauge invariant as we will explain below).

The contribution from a spectator with $j^z=s^\prime$ to 
$\langle q ({\vec r}\times{\vec A})^z q \rangle$ thus
reads
\be
{\Delta}^z_{s^\prime} = -\frac{2}{3} \frac{g^2}{4\pi}
\int d^3r d^3r^\prime \psi_s^\dagger({\vec r})\psi_s({\vec r}) 
\psi_{s^\prime}^\dagger({\vec r}^\prime)
\frac{({\vec r}\times {\vec \alpha})^z}{|{\vec r}-{\vec r}^\prime|}
\psi_{s^\prime}({\vec r}^\prime) 
\ee
where the factor $-\frac{2}{3}$ arises from the color part of
the matrix element. Note that 
${\vec \Delta}_{s^\prime}$ is independent of the angular momentum
$j^z=s$ of the
`active quark', since 
$\psi_{\frac{1}{2}}^\dagger({\vec r})\psi_{\frac{1}{2}}({\vec r})
=\psi_{-\frac{1}{2}}^\dagger({\vec r})\psi_{-\frac{1}{2}}({\vec r})$.
However, it depends on the spin of the spectator since the 
orientation of the vector potential entering (\ref{eq:Delta})
depends on the latter.
For example, for $s^\prime=+\frac{1}{2}$, one finds
\be 
\psi_{s^\prime}^\dagger({\vec r}^\prime)
{({\vec r}\times {\vec \alpha})^z}
\psi_{s^\prime}({\vec r}^\prime)
= |{\cal N}|^2 j_0(kr^\prime)j_1(kr^\prime)2\frac{xx^\prime +
yy^\prime}{r^\prime}
\ee
and hence
\be
{\Delta}^z_{+\frac{1}{2}}= -\frac{2}{3}\alpha_s I_R
\ee
with
\be
I_R=|{\cal N}|^4 \int_{r<R} d^3r \int_{r^\prime<R} d^3r^\prime \left[j_0^2(kr)+j_1^2(kr)\right]2\frac{xx^\prime+yy^\prime}{|{\vec r}-{\vec r}^\prime|}
\frac{j_0(kr^\prime)j_1(kr^\prime)}{r^\prime}.
\label{eq:integral}
\ee
The angular integration can be easily done using the expansion
$\frac{1}{|{\vec r}-{\vec r}^\prime|} = \sum_{l=0}^\infty 
\frac{4\pi}{2l+1}
\sum_{m=-l}^l \frac{r_<^l}{r_>^{l+1}}Y_l^m(\Omega) 
Y_l^{m*}(\Omega^\prime)$ and noting that only $l=1$ contributes in
(\ref{eq:integral}), yielding
\be
I_R= \frac{4}{9}(4\pi)^2|{\cal N}|^4 \int_0^Rdr\int_0^Rdr^\prime {r^\prime}^2 r^3
\left[j_0^2(kr)+j_1^2(kr)\right] \frac{r_<}{r_>^2}
j_0(kr^\prime)j_1(kr^\prime)
= 0.1165
\ee
which does not depend on the bag radius.
For $s^\prime = -\frac{1}{2}$ the result is identical with a negative
sign.

If the active quark has $s$ alligned with that of the proton
the two spectators must have opposite $s^\prime$ and their
contribution to ${\vec \Delta}$ cancels, i.e. ${\vec \Delta}$ is
nonzero only in those wave function components where the active quark
has $s=-\frac{1}{2}$, in which case both spectators have
$s^\prime=+\frac{1}{2}$. As a result, $\Delta^z_{q/p}$ is equal to
twice $\Delta^z_{+\frac{1}{2}}$ times the probability to find
that quark flavor with $s=-\frac{1}{2}$ (which is $\frac{1}{3}$ for
$q=u$ and $\frac{2}{3}$ for $q=d$, and hence
\bea
\label{eq:main}
\Delta_{u/p}^z &=& \frac{2}{3}{\Delta}^z_{+\frac{1}{2}}
=-\frac{4}{9} \alpha_s I_R=- 0.052\alpha_s\\
\Delta_{d/p}^z &=& \frac{4}{3}{\Delta}^z_{+\frac{1}{2}}
=-\frac{8}{9} \alpha_s I_R=
- 0.104\alpha_s \nonumber
\eea
which is the main result of this paper.

For simplicity, we evaluated the vector potential in the 
${\vec \nabla}\cdot {\vec A}=0$ gauge. However, 
our perturbative 
result is gauge invariant (to ${\cal O}(\alpha_s)$) at least
in the subclass of all gauges where all color components are
treated (globally) SU(3)-symmetrically. I such gauges, matrix elements
of operators of the type $q^\dagger \lambda^a \Gamma q A^a$,
where $\Gamma$ is some Dirac matrix, and $A^a$ is calculated 
to ${\cal O}(\alpha_s)$, are proportional to the
matrix elements of the corresponding abelian operators.
Therefore it is sufficient to establish gauge invariance of 
$q^\dagger {\vec r}\times {\vec A} q$ for abelian fields.
The key observation is that the MIT bag-model wave
functions contain no correlations between the positions of the 
quarks. Therefore, after eliminating the 
color in this calculation and introducing abelian currents, 
$\Delta_{s^\prime}^z$ factorizes into the
density of the active quark $\psi_s^\dagger({\vec r})\psi_s({\vec r})$
times $({\vec r}\times {\vec A})_z = rA_\phi$. 
Writing the volume integral $\int d^3r$ in cylindrical coordinates, 
one can isolate the only $\phi$-dependent term $r\int_0^{2\pi}d\phi
A_\phi= \oint d{\vec r}\cdot {\vec A}({\vec r})$ as a closed loop
integral with fixed $r$ and $z$. The closed loop integral is 
gauge invariant (its numerical value represents the color-magnetic
flux through a circle with radius $r$) and so is the volume
integral in which it enters.

In addition to (\ref{eq:JJi}) and (\ref{eq:JJM}), other 
angular momentum decompositions have been proposed in the literature
(see for example Refs. \cite{CG,Wakamatsu}). Since we used Coulomb 
gauge, ${\vec A}=
{\vec A}_{phys}$, where ${\vec A}_{phys}$ is the transverse part of 
the vector potential as defined in Refs. \cite{CG,Wakamatsu}.
In the decomposition proposed in Ref. \cite{CG}, the quark orbital 
angular momentum involves only the longitudinal part of the
vector potential, which is zero in our calculation, and which, in
the bag model, as well as many other quark models, 
vanishes also in other gauges to ${\cal O}(\alpha_s)$ as discussed
above. Therefore, within our approximations, the quark orbital 
angular momentum as introduced in Ref. \cite{CG} would not be affected
by spectator effects and would be given by Eq. (\ref{eq:num1}).
In contradistinction, the quark orbital 
angular momentum in Ref. \cite{Wakamatsu} is identical
to that of \cite{JiPRL} and therefore would receive the same 
contribution from the vector potential as $L_q^z$.

\section{Discussion}
We found that the effect from the vector potential caused by
spectators on 
$\Delta_{q/p}^z\equiv L_q^z-{\cal L}_q^z$ is negative
for both $q=u,d$. This is consistent with the sign that one may have 
intuitively expected: since ${\vec A}$ is obtained using the 
Biot-Savard equation ${\vec A}({\vec r})=\frac{1}{4\pi}
\int d^3r^\prime 
\frac{{\vec j}({\vec r})}{|{\vec r}-{\vec r}^\prime|}$, one would
expect that ${\vec A}$ and ${\vec j}$ have the same orientation.
Hence ${\vec r}\times{\vec A}$ should have the same sign as the
magnetization density ${\vec r}\times{\vec j}$. 
In the sum over all spectators, quarks with 
$j_z=+\frac{1}{2}$ dominate and thus the magnetization density of
the spectators has the same sign as the nucleon spin.
Taking into 
consideration the negative color factor which reflects the fact that
the interaction between the quarks is attractive, and the fact
that only spectators with $s^\prime$ contribute, one arrives at
the above negative sign.

The difference $\Delta_{u/p}^z-\Delta_{d/p}^z = 0.052\alpha_s$ is 
positive. This implies that the vector potential from the
spectators  adds a negative contribution to ${\cal L}_u^z - {\cal L}_d^z$ 
when compared to $L_u^z-L_d^z$. Since $L_u^z-L_d^z$ calculated in 
lattice QCD is already negative, subtracting our result for
$\Delta_{u/p}^z-\Delta_{d/p}^z$ would make it even more negative, and
therefore
this does not help in understanding the puzzling lattice results.

The numerical values of $\Delta_{q/p}^z$ are of the same order of 
magnitude as the corresponding difference for a free quark 
\cite{BC}.
Whether our result is overall numerically significant depends on the numerical
value of $\alpha_s$ --- a matter of debate in the context of
the bag model. For example, in order to generate the experimentally
measured $N-\Delta$ mass splittings from perturbative color-hyperfine
interactions in the bag model,  
$\alpha_s={\cal O}(1)$ is required --- in which case the 
$\Delta_{q/p}^z$
would be large enough that $L_{d/p}$ would become positive,
but still not large enough to render $L_{u/p}$ negative.
Of course, such a large value of $\alpha_s$ is inconsistent with the
use of perturbation theory, and may only reflect the need for
correlations between quark wave functions, which are not present
in the MIT bag model. However, nonperturbative correlations that 
enhance the hyperfine splitting, may also enhance $\Delta_q^z$.
Since both $\Delta_u^z$ and $\Delta_d^z$ are negative, their 
difference is small which seems to indicate that spectator 
contributions to the isovector combination $L_u-L_d$ are small, but
may not be small in the case of the isoscalar combination
$L_u+L_d$.

{\bf Acknowledgements:}
This work was supported by the DOE under grant number 
DE-FG03-95ER40965.  

\bibliography{oam_bag.bbl}

\begin{thebibliography}{3}
\bibitem{EMC} J. Ashman et al (EMC), Phys. Lett. B {\bf 206}, 364
(1988); Nucl. Phys. B {\bf 328}, 1 (1989).
\bibitem{JiPRL} X. Ji, Phys. Rev. Lett. {\bf 78}, 610 (1997).
\bibitem{lattice} Ph. H\"agler {\it et al.} (LHPC Collaboration),
Phys. Rev. D {\bf 77}, 094502 (2008).
\bibitem{Tony} A.W.~Thomas, and F. Myhrer, Phys. Lett. B {\bf 663}, 
302 (2008); A.W.~Thomas, Phys. Rev. Lett. {\bf 101}, 102003 (2008).
\bibitem{JM} R.L. Jaffe and A. Manohar, Nucl. Phys. {\bf B337}, 509
(1990).
\bibitem{MIT} A. Chodos et al., Phys. Rev. D{\bf 9},
3471 (1974); T.A. DeGrand et al., {\sl ibid.} {\bf 12} 2060 (1975).
\bibitem{BC} M. Burkardt and H. BC, Phys. Rev. D{\bf 79}, 071501
(2009).
\bibitem{CG} X.S. Chen et al., Phys. Rev. Lett. {\bf 100}, 232002
(2008); {\bf 103}, 062001 (2009).
\bibitem{Wakamatsu} M. Wakamatsu, Phys. Rev. D
{\bf 81}, 114010 (2010); Eur. Phys. J. A {\bf 44}, 297 (2010).
\end{thebibliography}
\end{document}